\documentclass{ws-procs9x6}

\begin{document}

\title{Problems with the MSSM :\\
mu \& proton decay}

\author{S. Raby$^*$}

\address{Department of Physics, The Ohio State University,\\
191 W. Woodruff Ave, Columbus, OH 43210, USA\\
$^*$E-mail: raby@mps.ohio-state.edu}

\begin{abstract}
Contribution to KMI Inauguration Conference "Quest for the Origin of Particles and the  Universe" (KMIIN), 24-26 Nov. 2011, KMI, Nagoya University.

\end{abstract}


\bodymatter

\section{Review the problems \& proposed solutions in the literature
}

The most general $SU(3)_C \times SU(2)_L \times U(1)_Y$  invariant
superpotential up to operators of dimension 5 is given by,
\begin{eqnarray} W = & Y_e^{i j} H_d L_i \bar E_j + Y_d^{i j} H_d Q_i \bar D_j + Y_u^{i j} H_u Q_i \bar U_j & \\
&  + \mu H_u H_d +  \kappa_{i j}^{(0)} H_u L_i H_u L_j & \\
&  +  \kappa_i L_i H_u + \lambda_{i j k}^{(0)} L_i L_j \bar E_k + \lambda_{i j k}^{(1)} L_i Q_j \bar D_k + \lambda_{i j k}^{(2)} \bar U_i \bar U_j \bar D_k & \\
& +  \kappa_{i j k l}^{(1)} Q_i Q_j Q_k L_k  + \kappa_{i j k l}^{(2)} \bar U_i \bar U_j \bar D_k \bar E_k  + \kappa_{i j k}^{(3)} Q_i Q_j Q_k {H_d}  & \\
& +  \kappa_{i j k}^{(4)} Q_i \bar U_j \bar E_k {H_d} + \kappa_i^{(5)} L_i H_u H_u H_d &
\end{eqnarray}
Let us now discuss this superpotential term by term.

Equations 1 and 2 are terms which are absolutely necessary in the effective low energy theory.  Eqn. 1 includes the Yukawa couplings for quark and charged lepton masses.   While eqn. 2 contains the so-called $\mu$ term, necessary to avoid unwanted massless states, and the Weinberg neutrino mass operator.  The terms in equations 3, 4 and 5 are not acceptable in the low energy theory.    Eqn. 3 contains the R-parity violating operators.   In any SUSY GUT, such as $SU(5)$, the last three terms are all generated by the $SU(5)$ invariant operator
\begin{equation}  10 \bar 5 \bar 5 \supset  L L \bar E + L Q \bar D + \bar U \bar D \bar D .
\end{equation}
There are strong constraints from the non-observation of proton decay on the product  $\lambda^{(1)} \lambda^{(2)}$.
\begin{equation} \tau(p \rightarrow e^+ \pi^0) > 10^{34} {\rm years} \end{equation}  gives
\begin{equation}
\lambda^{(1)} \lambda^{(2)} \leq 10^{-27} .  \label{eq:lambda}
\end{equation}
Equations 4 and 5 are dimension 5 baryon and lepton number violating operators.   The strong bound on the proton lifetime
\begin{equation} \tau(p \rightarrow K^+ \bar \nu) > 2.3 \times 10^{33} {\rm years} \end{equation} gives the significant bounds on the
$\kappa$ parameters.   In particular,  \begin{equation} \kappa^{1, 2} \equiv \frac{1}{\Lambda}  \leq  10^{-27} {\rm GeV}^{-1} . \end{equation}

\section{Discrete  non-R symmetries \& mu
}
In this section we consider some discrete non-R symmetries which were invented to address problems of
nucleon decay due to dimension 4 and/or 5 operators.  The list of such symmetries includes  R parity \cite{Farrar:1978xj}/matter parity \cite{Dimopoulos:1981dw} [$M_2$], baryon triality  \cite{Ibanez:1991pr} [$B_3$] and proton hexality  \cite{Dreiner:2005rd} [$P_6$].
All three of these discrete symmetries are anomaly free.  The states in the minimal supersymmetric standard model [MSSM] have charges given
in Table \ref{aba:nonR}.
\begin{center}
\begin{table}
\tbl{}
{\tablefont
\begin{tabular}{llllllll}
\toprule   & $Q$ & $\bar U$ & $\bar D$ & $L$ & $\bar E$ &  $H_u$ &  $H_d$ \\\colrule
$M_2$ & 1  & 1 & 1 & 1 & 1 & 0 & 0 \\
$B_3$ &  0 & 2 & 1 & 2 & 2  & 1 & 2 \\
$P_6$ & 0  & 1 & 5 & 4 & 1 & 5 & 1 \\\botrule
\end{tabular}}\label{aba:nonR}
\end{table}
\end{center}

The MSSM is consistent with gauge coupling unification and thus grand unification.
Note, however, that neither baryon triality, $B_3$, nor proton hexality, $P_6$, charges are consistent with grand unification.   Nor can they be
made consistent by adding some linear combination of hypercharge.  Only matter parity, $M_2$, is consistent with grand unification, since it
gives the same charge to all fields in both the $10 \supset \{ Q, \bar U, \bar E \}$ and $\bar 5 \supset \{ \bar D, L \}$ of $SU(5)$.  The Higgs
fields don't however come in complete $SU(5)$ multiplets,  so they must be treated special.   We will come back to this issue later.

Consider the possible MSSM operators up to order dimension 5.  In Table \ref{aba:nonR_operators} we show how they transform under the discrete non-R symmetries.   Note that only operators with total charge zero are allowed in the superpotential.   The operators in the upper three lines are all allowed.  These include all Yukawa couplings, the $\mu$ term and the Weinberg neutrino mass operator.  These are all operators we need in the effective low energy theory.  The next set of operators all violate matter parity and proton hexality.  Baryon triality, on the other hand, permits all the lepton number violating operators, while forbidding the baryon number violating operators.   Thus all three symmetries prevent rapid proton decay.   In addition, the lightest supersymmetric particle [LSP] is stable in the case of matter parity and proton hexality,  but not for baryon triality.  Finally matter parity allows all dimension 5 operators, while they are forbidden in the case of baryon triality and proton hexality.
\begin{center}
\begin{table}
\tbl{}
{\tablefont
\begin{tabular}{llll}
\toprule $\mathbb{Z}_M$ & $M_2$ & $B_3$ & $P_6$ \\\colrule
$Q \bar U H_u$, ... & 0  & 0 & 0 \\
$H_u H_d$  &  0 & 0 & 0  \\
$(L H_u)^2$ & 0  & 0 & 0  \\\colrule
$\bar U \bar D \bar D$ & 1  & 1 & 5  \\
$Q L \bar D$ & 1  & 0 & 3  \\
$L L \bar E$ & 1  & 0 & 3  \\
$H_u L$ & 1  & 0 & 3  \\\colrule
$Q Q Q L$ & 0  & 2 & 4  \\
$\bar U \bar U \bar D \bar E$ & 0  & 1 & 2  \\\botrule
\end{tabular}}\label{aba:nonR_operators}
\end{table}
\end{center}
To summarize, what would we like to do ?
\begin{itemlist}[(d)]
\item Forbid the $\mu$ term perturbatively.
\item Forbid  dimension  3, 4 \&  5  baryon and lepton number violating operators.
\item And do this with a discrete symmetry which is consistent with grand unification for quarks and leptons!
\end{itemlist}

Let us now discuss anomaly cancelation in more detail.
Below we give the anomaly coefficients, where  $A_i$ is the coefficient for the $SU(i)\times SU(i) \times \mathbb{Z}_M$ non-R symmetry.
\begin{eqnarray}
A_3 = & \frac{1}{2} \sum_{g = 1}^3 (3 \cdot q_{10}^g + q_{\bar 5}^g) & \\
A_2 = & \frac{1}{2} \sum_{g = 1}^3 (3 \cdot q_{10}^g + q_{\bar 5}^g)  +  \frac{1}{2} (q_{H_u} + q_{H_d}) & \\
A_1 = & \frac{1}{2} \sum_{g = 1}^3 (3 \cdot q_{10}^g + q_{\bar 5}^g)  +  \frac{3}{5} \cdot \frac{1}{2} (q_{H_u} + q_{H_d}) &
\end{eqnarray}
In order to be anomaly free, these coefficients must satisfy
\begin{equation} (A_{1 \leq i \leq 3} \ {\rm mod} \ \eta) = \frac{1}{24} (A_0 \ {\rm mod} \ \eta) = \rho \end{equation} with
\begin{equation} \eta = [ M \; {\rm for} \  M \ {\rm odd};\;\;  M/2 \; {\rm for} \ M \ {\rm even} ].   \end{equation}
When $\rho = 0$ the theory is anomaly free.  However, when $\rho \neq 0$ the anomalies are absent due to Green-Schwarz anomaly
cancelation.

We now want to show that a $\mu$-term cannot be forbidden by a discrete non-R symmetry.  The proof follows by contradiction.  Anomaly cancelation
requires  \begin{equation} A_2 - A_3 = 0 \, {\rm mod} \ \eta \end{equation}  or  \begin{equation} \frac{1}{2} (q_{H_u} + q_{H_d}) = 0  \, {\rm mod} \ \eta. \end{equation}  However, a $\mu$-term is allowed when \begin{equation} (q_{H_u} + q_{H_d}) = 0 \;  {\rm mod} \ M . \end{equation}   Hence a $\mu$-term is always consistent with anomaly cancelation of a non-R symmetry.

To summarize,
\begin{itemlist}[(d)]
\item $M_2$ :  eliminates  Dim 3 \& 4,   B \& L violating operators;  LSP stable
\item  $B_3$ :  eliminates  Dim. 4,  B and Dim. 5,  B \& L violating operators; LSP unstable
\item $P_6$  :  eliminates  Dim. 3, 4 \& 5, B \& L violating operators;  LSP stable
\item  The MSSM is consistent with GUTs,  BUT  $B_3$ \& $P_6$ are NOT!
\item  The $\mu$-term is allowed by ALL
\end{itemlist}

\section{Discrete R symmetries $\mathbb{Z}^R_M$ }

Let us now consider discrete R symmetries.  In particular, we will focus on a simple $\mathbb{Z}^R_4$ symmetry.  This section is based on the analysis in the following papers \cite{Lee:2010gv,Lee:2011dya,Kappl:2010yu}.  An earlier paper \cite{Babu:2002tx} also found the same discrete R symmetry.
Consider the discrete $\mathbb{Z}^R_M$ R symmetry.  Under this symmetry,  chiral superfields with charge $q$ transform with the phase
\begin{equation}  (\gamma_M)^q =  \exp(\frac{2 \pi i q}{M}). \end{equation}   For charge assignments consistent with an $SU(5)$ GUT, we have
\begin{equation}  q =  \{ q_{10}, q_{\bar 5}, q_{H_u}, q_{H_d} \} \label{eq:superfields} \end{equation} for chiral superfields transforming in the $ \{ 10, \bar 5 \}$ representations
of $SU(5)$.  This includes the three families of quarks and leptons.   However the Higgs multiplets come in incomplete $SU(5)$ representations and thus
are treated separately.  We will show later that this is consistent with orbifold GUTs in more than three spacial dimensions or heterotic string constructions.
The charges in Eqn. (\ref{eq:superfields}) are for the bosons in the chiral superfields.  The chiral fermions, however, transform with charges $q - 1$.  With this convention,  gauginos have charge, +1, and the superpotential necessarily transforms with charge, +2.

Given these new charges for the fermions,  the anomaly coefficients are now given by
\begin{eqnarray}
A_3^R = & \frac{1}{2} \sum_{g = 1}^3 (3 \cdot q_{10}^g + q_{\bar 5}^g) - 3 & \\
A_2^R = & \frac{1}{2} \sum_{g = 1}^3 (3 \cdot q_{10}^g + q_{\bar 5}^g)  +  \frac{1}{2} (q_{H_u} + q_{H_d}) - 5 & \\
A_1^R = & \frac{1}{2} \sum_{g = 1}^3 (3 \cdot q_{10}^g + q_{\bar 5}^g)  +  \frac{3}{5} \cdot [\frac{1}{2} (q_{H_u} + q_{H_d}) - 11 ] &
\end{eqnarray}

Canceling anomalies then requires
\begin{eqnarray} A_2^R - A_3^R = 0 \; {\rm mod} \ \eta & \; {\rm or}  \;\; (q_{H_u} + q_{H_d}) = 4  \; {\rm mod} \ 2 \eta & \label{eq:mu} \\
 A_1^R - A_3^R = 0 \; {\rm mod} \ \eta & \; {\rm or} \;\;   \frac{3}{5} \cdot [\frac{1}{2} (q_{H_u} + q_{H_d}) - 6 ] = 0  \; {\rm mod} \ \eta. & \label{eq:mu2} \end{eqnarray}
 The general solution to these equations is given by \begin{equation}  \mathbb{Z}_M^R  \;\; {\rm with} \;\; M = 3, 4, 6, 8, 12 \ {\rm or} \; 24. \end{equation}
Note,  since a $\mu$-term must satisfy the condition  \begin{equation} (q_{H_u} + q_{H_d}) = 2 \;  {\rm mod} \ M, \end{equation} it is forbidden for any $M \neq 2$.

It is easy to show that the baryon and lepton number violating dimension 5 operators $Q Q Q L$ and $\bar U \bar U \bar D \bar E$ are also forbidden.  Consider the constraint equation which follows from allowing Yukawa couplings for both up and
down quarks.   We have \begin{equation} 3 q_{10} + q_{\bar 5} + q_{H_u} + q_{H_d}  = 4 \; {\rm mod} \ M. \end{equation}  However, using the result of Eqn. (\ref{eq:mu}) we have \begin{equation} 3 q_{10} + q_{\bar 5} = 0, \end{equation} which is inconsistent with having the terms, $Q Q Q L$ and $\bar U \bar U \bar D \bar E$, in the superpotential.

In Ref. \cite{Lee:2011dya} we searched for all possible discrete R symmetry charge assignments to the anomaly constraints in Eqns. (\ref{eq:mu}, \ref{eq:mu2}) satisfying
\begin{enumerate}
\item $\mu$-term is forbidden;
\item mixed gauge-$\mathbb{Z}_M^R$ anomalies cancel, i.e. $A_{1 \leq i \leq 3}^R = \rho \; {\rm mod} \ \eta$;
\item Yukawa couplings,  $10 \ 10 \ H_u$, $10 \ \bar 5 \ H_d$ as well as the neutrino mass Weinberg operator $\bar 5 \ H_u \ \bar 5 \ H_u $ are allowed;
\item R parity violating couplings are forbidden.
\end{enumerate}

The solutions are given in Table \ref{aba:solutions}.
\begin{center}
\begin{table}
\tbl{}
{\tablefont
\begin{tabular}{llllll}
\toprule $M$ & $q_{10}$ & $q_{\bar 5}$ & $q_{H_u}$ & $q_{H_d}$ &  $\rho$ \\\colrule
4 & 1  & 1 & 0 & 0 & 1 \\
6  &  5 & 3 & 4 & 0 & 0 \\
8 & 1  & 5 & 0 & 4 & 1  \\
12 & 5  & 9 & 4 & 0 & 3  \\
24& 5  & 9 & 16 & 12 & 9  \\\botrule
\end{tabular}}\label{aba:solutions}
\end{table}
\end{center}

\section{Unique $\mathbb{Z}^R_4$}

Consider now the unique discrete R symmetry consistent with $SO(10)$, $\mathbb{Z}_4^R$.   From Table \ref{aba:solutions} we obtain
the charge assignments in Tables \ref{aba:z4r} and \ref{aba:z4r2}.

\begin{center}
\begin{table}
\tbl{$SU(5)$}
{\tablefont
\begin{tabular}{lllll}
\toprule $\mathbb{Z}_4^R$ & $q_{10}$ & $q_{\bar 5}$ & $q_{H_u}$ & $q_{H_d}$  \\\colrule
 & 1 & 1  & 0 & 0  \\\botrule
\end{tabular}}\label{aba:z4r}
\end{table}
\end{center}
\begin{center}
\begin{table}
\tbl{$SO(10)$}
{\tablefont
\begin{tabular}{lll}
\toprule $\mathbb{Z}_4^R$ & $q_{16} = q_{10} = q_{\bar 5} = q_1$ & $q_{10} = q_{H_u} = q_{H_d}$  \\\colrule
 & 1  & 0  \\\botrule
\end{tabular}}\label{aba:z4r2}
\end{table}
\end{center}
Given these charges the possible terms in the superpotential are given in Table \ref{aba:allowed_z4r}, along with their charge under $\mathbb{Z}_4^R$.  Note
only terms with charge +2 are allowed in the superpotential.   The first line in Table \ref{aba:allowed_z4r} includes all Yukawa-type coupling terms.
\begin{center}
\begin{table}
\tbl{Allowed terms in the superpotential consistent with $\mathbb{Z}_4^R$ have charge +2.}
{\tablefont
\begin{tabular}{ll}
\toprule Operator & $\mathbb{Z}_4^R$ charge    \\\colrule
$Q \ \bar U \ H_u, \dots$ &  2 \\
$H_u \ H_d$ & 0    \\
$(L \ H_u)^2$ &  2  \\
$\bar U \ \bar D \ \bar D$ & 3 \\
$Q \ L \ \bar D$ & 3 \\
$L \ L \ \bar E$ & 3 \\
$H_u \ L$ &  1 \\
$Q \ Q \ Q \ L$ & 0 \\
$\bar U \ \bar U \ \bar D \ \bar E$ & 0 \\
$W$ & 2 \\ \botrule
\end{tabular}}\label{aba:allowed_z4r}
\end{table}
\end{center}

Thus the allowed perturbative superpotential consistent with $\mathbb{Z}_4^R$ is given by
\begin{equation}  W_p =  Y_e^{i j} H_d L_i \bar E_j + Y_d^{i j} H_d Q_i \bar D_j + Y_u^{i j} H_u Q_i \bar U_j +  \kappa_{i j}^{(0)} H_u L_i H_u L_j. \end{equation}  In general the full superpotential, $W$,  may include non-perturbative contributions, i.e. $W = W_p + \Delta W_{non-perturbative}$.
Note, non-perturbative contributions to the superpotential can come from gaugino condensates, instantons, etc.   All such contributions are of the form \begin{equation}  \Delta W_{np} \supset \langle W \rangle_0 \cdot ( \dots ), \end{equation} since the non-perturbative vacuum expectation value
$ \langle W \rangle_0$ has charge, +2, under $\mathbb{Z}_4^R$.   Thus, in general, we obtain
\begin{equation} \Delta W_{np} = \exp \left(-8 \pi^2 \frac{1 + 2 n}{1 + 2 \nu} S \right) \left[ B_0 + \bar \mu \ H_u  H_d +
\bar \kappa_{i j k l}^{(1)} Q_i Q_j Q_k L_k  + \bar \kappa_{i j k l}^{(2)} \bar U_i \bar U_j \bar D_k \bar E_k \right], \end{equation}  i.e.
where all terms after the pre-factor, $ \langle W \rangle_0$,  have charge, 0.  Under the discrete R symmetry $S \rightarrow S + \frac{i}{2} \Delta_{GS}$, where $\Delta_{GS}$ is the Green-Schwarz phase.  It is this transformation which guarantees that $\langle W \rangle_0$ has charge, +2, under $\mathbb{Z}_4^R$.  It is important to realize that non-perturbative effects do not break matter parity, which is a subgroup of $\mathbb{Z}_4^R$. Thus matter parity is a good symmetry of the low energy theory.

How large can we expect these non-perturbative corrections to be?   In gravity-mediated supersymmetry breaking we have \begin{equation} m_{3/2} \sim
\langle W \rangle_0/M_{Pl}^2. \end{equation}  Hence we expect from dimensional analysis
\begin{equation}   \Delta W_{np} \propto   B_0 \ m_{3/2} M_{Pl}^2  + m_{3/2} \ H_u  H_d +
\frac{m_{3/2}}{M_{Pl}^2} \ [ Q Q Q L  +  \bar U \bar U \bar D \bar E ].\end{equation}  Thus we obtain  \begin{eqnarray}  & \mu \sim m_{3/2} & \\
& 1/\Lambda \sim \frac{m_{3/2}}{M_{Pl}^2} & \approx 10^{-33} \; {\rm GeV}^{-1}. \label{eq:lambda2} \end{eqnarray}  By comparing Eqns. (\ref{eq:lambda} and \ref{eq:lambda2}) we see that dimension 5 baryon and lepton number violating operators are so suppressed that they are unlikely to produce observable effects.  To summarize, {\em the discrete $\mathbb{Z}_4^R$ symmetry defines the MSSM.}

\subsection{Heterotic string construction \cite{Kappl:2010yu}}
\begin{figure}
\begin{center}
\epsfig{file=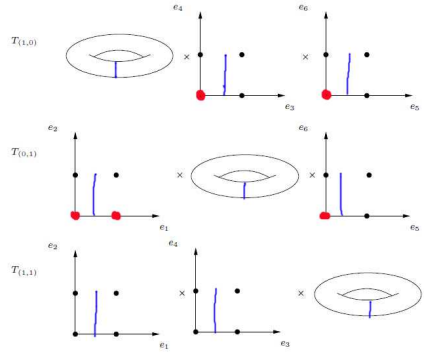,width=2.5in}
\end{center}
\caption{$E(8) \times E(8)$ heterotic string compactified on this 6D internal orbifold, given by
$(T^2)^3/(\mathbb{Z}_2 \times \mathbb{Z}_2)$ plus a shift by half a unit in the upward direction.}
\label{aba:fig1}
\end{figure}

Let us now discuss an example of the $\mathbb{Z}_4^R$ symmetry obtained in an orbifold construction of the $E(8) \times E(8)$ heterotic
string compactified on a 6 dimensional internal orbifold, given by $(T^2)^3/(\mathbb{Z}_2 \times \mathbb{Z}_2)$ plus a shift by half a unit in the upward direction (see Fig. (\ref{aba:fig1})) \cite{Kappl:2010yu}.   The orbifold has a non-trivial fundamental group,  $\pi(1) = \mathbb{Z}_2$.  And there is a $\mathbb{Z}_2$ Wilson line described by the blue lines in Fig. (\ref{aba:fig1}).  The red points show where the three families of quarks
and leptons sit, in complete $SO(10)$ spinor representations.
As a consequence of the $\mathbb{Z}_4^R$ symmetry the $E(8) \times E(8)$ heterotic string compactified on this
$(T^2)^3/(\mathbb{Z}_2 \times \mathbb{Z}_2)$ orbifold is very much like the MSSM.  The theory has the following properties.
\begin{itemlist}
\item It has the exact MSSM spectrum.
\item Supersymmetric vacua satisfying $F = D = 0$ are found which stabilize almost all moduli. At the same time,
the moduli vevs give mass to all vector-like exotics and extra gauge degrees of freedom.
\item  The $\mathbb{Z}_4^R$ symmetry eliminates or suppresses all dangerous
      baryon and lepton number violating operators.
\item  $\mu = 0$  perturbatively and  $\mu \sim m_{3/2}$ non-perturbatively.
\item $\mathbb{Z}_4^R$ comes partially from the Lorentz symmetry of the internal dimensions.
\item  The theory has a local $SO(10)$ gauge symmetry at certain fixed torii with quarks and leptons coming in complete spinor multiplets.
\item  The theory also has a $D(4)$ family symmetry with the lightest two families in a doublet and the third family and Higgs in a singlet
under the family symmetry.  Soft sequential breaking of the family symmetry can accommodate a family hierarchy and
suppress flavor changing neutral current processes \cite{Ko:2007dz}.
\item  It has gauge-top Yukawa coupling unification.  Both the third family and Higgs appear in the bulk of an effective $SU(6)$ orbifold GUT.
\item It has non-trivial rank 3 Yukawa matrices.
\item Lastly, the Higgs comes from chiral adjoint of an effective 6D SU(6) orbifold GUT.  As a consequence, it has been shown\cite{Brummer:2010fr}
that a $\mu$-term is generated proportional to $\langle W \rangle_0$.   This is due to the fact that the K\"{a}hler potential in this case is of the form \begin{equation} K \supset - \log \left( ( T + \bar T )( Z + \bar Z ) - ( H_u + \bar H_d )( \bar H_u + H_d ) \right). \end {equation}
\end{itemlist}
Note, a local $U(1)$ parity breaks the 6D orbifold GUT group $SU(6)$ to $SU(5)$.   Then the Wilson line breaks $SU(5)$ to the Standard Model gauge group. At the same time, the Wilson line gives mass to the color triplet Higgs fields, keeping the Higgs doublets massless.     In fact, only in orbifold
GUT theories is it possible to find a discrete R symmetry consistent with GUTs for quark and lepton families, while only the Higgs doublets are light. This result has been discussed in a recent paper \cite{Fallbacher:2011xg}.

\section{Singlet extensions of the MSSM}

Let me briefly discuss possible singlet extensions of the MSSM.

\subsection{GNMSSSM}

Consider first adding one new Standard Model singlet, $N$.  The generalized non-minimal supersymmetric standard model [GNMSSM] was then shown to be consistent with the $\mathbb{Z}_M^R$ symmetry with $M = 4, 8$ \cite{Lee:2011dya,Ross:2011xv}.   The $\mathbb{Z}_M^R$ charge assignments are given in Table \ref{aba:singlet}.
 \begin{center}
\begin{table}
\tbl{$SU(5)$}
{\tablefont
\begin{tabular}{llllll}
\toprule $M$ & $q_{10}$ & $q_{\bar 5}$ & $q_{H_u}$ & $q_{H_d}$ & $q_N$ \\\colrule
4 & 1 & 1  & 0 & 0 & 2 \\
8 & 1 & 5 & 0 & 4 & 6 \\\botrule
\end{tabular}}\label{aba:singlet}
\end{table}
\end{center}
The superpotential is given by $ W = W_p + \Delta W_{np} $ with the perturbative part-  \begin{equation}  W_p = W_{MSSM}^{\mu = 0} + \lambda \ N \ H_u  H_d
+ \kappa \ N^3 .\end{equation}   And possible non-perturbative contributions given by-
\begin{equation} \Delta W_{np} \sim m_{3/2} M_{Pl}^2 + m_{3/2}^2 \ N + m_{3/2} \ N^2 + m_{3/2} \ H_u  H_d . \end{equation}  Note, this model
does not have cosmological problems associated with domain walls.

\subsection{Axion compatible solution}

The symmetry $\mathbb{Z}_{24}^R$ allows for a possible axion solution to the strong CP problem \cite{Lee:2011dya}.  A simple
model has two additional Standard Model singlets,  $N,\ X$, with charges given by
 \begin{center}
\begin{table}
\tbl{$SU(5)$}
{\tablefont
\begin{tabular}{llllll}
\toprule  $q_{10}$ & $q_{\bar 5}$ & $q_{H_u}$ & $q_{H_d}$ & $q_N$ & $q_X$ \\\colrule
5 & 9 & 16  & 12 & -1 & 5 \\\botrule
\end{tabular}}\label{aba:singlet2}
\end{table}
\end{center}
and superpotential \begin{equation}  W = \frac{\alpha}{M_{Pl}} \ N^2 \ H_u  H_d + \frac{\beta}{M_{Pl}} \ X \ N^3. \end{equation}

The second term is needed to generate a vacuum expectation value, $\langle N \rangle \sim 10^{10 - 11}$ GeV which then results in a
$\mu$-term with $\mu \sim 10^{2 - 3}$ GeV.   Although it is not as obvious, but the symmetry $\mathbb{Z}_{24}^R$ is the only one
for which non-perturbative contributions to the axion  mass are small compared to the standard QCD contribution to its mass.

\section{Conclusions}

In conclusion, we have found the unique $\mathbb{Z}_4^R$ symmetry which forbids dimension 3, 4 \& 5 operators to all orders
in perturbation theory.  The $\mu$-term is generated non-perturbatively of order the weak scale.
We also found a heterotic orbifold realization of the $\mathbb{Z}_4^R$ symmetry.
In general we showed that there are $\mathbb{Z}_{4,6,8,12,24}^R$ symmetries consistent with $SU(5)$ which also accomplish the above
results.  In an important sense, these discrete R symmetries define the MSSM.  Lastly we discussed models with additional singlets.  This included a generalized NMSSM defined by a $\mathbb{Z}_{4, 8}^R$ symmetry.  We also considered a model which allows an axion solution to the strong CP problem defined by a $\mathbb{Z}_{24}^R$ symmetry.

\end{document}